\documentclass[conference]{IEEEtran}
\pdfoutput=1
\usepackage[cmex10]{amsmath} 
\usepackage{amssymb,mathrsfs}
\usepackage{pgf,tikz}
\usepackage{epsfig}
\usepackage{citesort}
\usepackage{url}
\usetikzlibrary{arrows}

\usepackage{mdwmath,mdwtab}
\usepackage[font=footnotesize,caption=false]{subfig}

\newtheorem{theorem}{Theorem}

\newtheorem{proof}{Proof}

\newtheorem{definition}{Definition}

\newtheorem{corollary}{Corollary}

\title{Distributed Consensus with Finite Messaging}

\author{
  \IEEEauthorblockN{Debashis Dash and Ashutosh Sabharwal}
  \IEEEauthorblockA{Department of Electrical and Computer Engineering\\
    Rice University\\
    Houston, TX 77005\\
    Email: \{ddash, ashu\}@rice.edu
  }
}

\begin{document}
\maketitle
\begin{abstract}
Inspired by distributed resource allocation problems in dynamic topology networks, we initiate the study of distributed consensus with finite messaging passing. We first find a sufficient condition on the network graph for which no distributed protocol can guarantee a conflict-free allocation after $R$ rounds of message passing. Secondly we fully characterize the conflict minimizing zero-round protocol for path graphs, namely random allocation, which partitions the graph into small conflict groups. Thirdly, we enumerate all one-round protocols for path graphs and show that the best one further partitions each of the smaller groups. Finally, we show that the number of conflicts decrease to zero as the number of available resources increase.
\end{abstract}

\section{Introduction}
One of the important problems in distributed wireless networks is allocating system resources (e.g. frequency bands or time-slots etc.) in a decentralized fashion, see e.g~\cite{ram99,ca06,xln06}.  
 Typically nodes need global network state information to achieve optimal allocation. Due to many practical constraints, e.g. mobility of nodes and overhead in obtaining side information etc. such global knowledge is usually not available to all the nodes. For such scenarios, local message passing, which builds this network knowledge either implicitly or explicitly has been shown to be robust for achieving distributed consensus \cite{bgp06}. In most cases such robustness is achieved asymptotically \cite{mbk09,bgp06}. 
However, if optimality is unattainable with local information, the key question then becomes, what are the best distributed protocols and how far are these protocols from the optimal allocation \cite{lin92,pel00,aas10}, i.e. what is the performance loss due to distributed decisions based on finite rounds $(R)$ of message passing? Additionally, how the loss scales with available resources?

The problem of distributed allocation of orthogonal resources is related to the classical distributed graph coloring problem, i.e. coloring the nodes of a graph $G$ with $n$ nodes such that the connected nodes have different colors. If done with global information, a graph with maximum node degree $D$ can always be colored with $D+1$ colors using greedy schemes. For specific graphs, distributed schemes  require only $\Omega(\log n)$ rounds to achieve optimal coloring \cite{kk02,kos06,fgi09} but in general distributed coloring is not only NP hard \cite{kar72}, but it is hard to even approximately solve \cite{fk98,bgs98}. 
\begin{figure}[tbp]
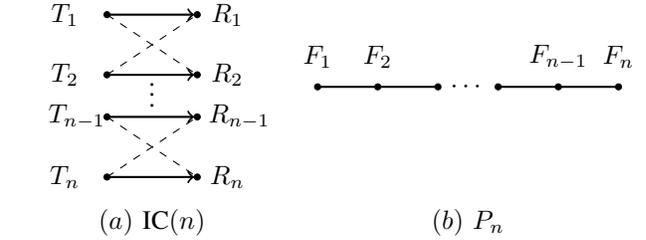

\centering
\tikz[scale=0.8]{
	\draw [fill=black] (0,0) circle (.05cm);
	\draw [fill=black] (1.5,0) circle (.05cm);
	\draw [fill=black] (0,1) circle (.05cm);
	\draw [fill=black] (1.5,1) circle (.05cm);
	\draw [fill=black] (0,1.7) circle (.05cm);
	\draw [fill=black] (1.5,1.7) circle (.05cm);
	\draw [fill=black] (0,2.7) circle (.05cm);
	\draw [fill=black] (1.5,2.7) circle (.05cm);
	\draw [->, thick] (0,0) -- (1.45,0);
	\draw [->, thick] (0,1) -- (1.45,1);
	\draw [->, thick] (0,1.7) -- (1.45,1.7);
	\draw [->, thick] (0,2.7) -- (1.45,2.7);
	\draw [dashed] (0,0) -- (1.5,1);
	\draw [dashed] (0,1) -- (1.5,0);
	\draw (0.75,1.5) node {$\vdots$};
	\draw [dashed] (0,2.7) -- (1.5,1.7);
	\draw [dashed] (0,1.7) -- (1.5,2.7);
	\draw (-0.7,0) node {$T_n$};
	\draw (2,0) node {$R_n$};
	\draw (-.5,1) node {$T_{n-1}$};
	\draw (2.2,1) node {$R_{n-1}$};
	\draw (-.7,1.7) node {$T_2$};
	\draw (2,1.7) node {$R_2$};
	\draw (-.7,2.7) node {$T_1$};
	\draw (2,2.7) node {$R_1$};
	\draw (0.75,-0.75) node {$(a)~\text{IC}(n)$};
	\draw [fill=black] (3.5,1.5) circle (.05cm);
	\draw [fill=black] (4.5,1.5) circle (.05cm);
	\draw [fill=black] (5.5,1.5) circle (.05cm);
	\draw [fill=black] (6.5,1.5) circle (.05cm);
	\draw [fill=black] (7.5,1.5) circle (.05cm);
	\draw [fill=black] (8.5,1.5) circle (.05cm);
	\draw [thick] (3.5,1.5) -- (4.5,1.5);
	\draw [thick] (4.5,1.5) -- (5.5,1.5);
	\draw (6,1.5) node {$\cdots$};
	\draw [thick] (6.5,1.5) -- (7.5,1.5);
	\draw [thick] (7.5,1.5) -- (8.5,1.5);
	\draw (3.5,2) node {$F_1$};
	\draw (4.5,2) node {$F_2$};
	\draw (7.5,2) node {$F_{n-1}$};
	\draw (8.5,2) node {$F_{n}$};
	\draw (6,-0.75) node {$(b)~P_{n}$};
}
\caption{(a) A cascaded $n$-user interference network with $n$ interfering flows. The intended communication links are shown by arrows and the interfering links are shown by dashed lines. (b) Flow graph of the network, where each node is represented as a single vertex, is a path network.
\label{fig:model}}
\vspace{-.5cm}
\end{figure}
%


We consider distributed vertex coloring with $c$ colors using only \emph{finite} number of messing rounds. Our contributions are four fold. Firstly we show that for $R$-hop symmetric networks (defined as networks which contain adjacent nodes that have identical $R$-hop neighborhoods), no distributed scheme can guarantee a proper coloring after $R$ rounds of message passing. This is because, to the nodes with symmetric neighborhood, the local network looks exactly the same. Hence any distributed scheme that relies on $R$-hop local information arrives at the same coloring for the adjacent nodes causing a defect, defined as an edge joining the same color nodes. Secondly, we show that for path networks $P_{n}$, shown in Figure~\ref{fig:model}, if the resources are randomly allocated, the network ends up with $O(\frac{n}{c})$ defects on an average. Random allocation partitions the network into smaller defect groups, defined as a connected subgraph with identically colored nodes. The defect distribution is proved to be proportional to the $n^{\text{th}}$ row of Pascal's triangle when $c=2$, which shows that typically the number of defects are neither too low nor too high, i.e. close to the average. Thirdly, we formulate all possible protocols based on the parameters that can be calculated by each node after one round of message passing. We calculate the defect distribution for an edge correcting protocol and a center correcting protocol and show that the best protocol resolves the defects inside each of the smaller groups keeping their edges fixed. Effectively, the random assignment breaks down the network into smaller defect groups and center correcting protocols do another round of random assignment inside the smaller defect groups.
Finally we show that all the protocols perform equally well and close to optimal as the number of colors increases. This is because random coloring produces very few defects for large $c$, thus an additional round provides very little benefit. 

The rest of the paper is organized as follows. In Section~\ref{sec:sysModel} we define the network, distributed protocols and the performance metric. In Section~\ref{sec:results} we present a converse for $R$-hop symmetric graphs and give achievable protocols for path graphs in Section~\ref{sec:achproto}. We finally conclude in Section~\ref{sec:conclusion}.

\section{System Model \label{sec:sysModel}}
\subsection{Network Description}
We consider networks with $n$ flows consisting of $n$ distinct transmitter-receiver pairs, each with a unique flow ID.
Orthogonal resource states, e.g. frequency bands or time-slots etc. are available that have to be distributedly allocated among the flows. 
The graph representation of the network is given by $G=(V,E)$, with $|V|=n$, where every flow is represented by a node $v\in V$ of the graph and the connected flows are represented by edges ($(i,j)\in E $ if node~$i$ and $j$ interfere with each other). $G$ is called the \textit{flow graph} of the network (just graph, henceforth). The orthogonal resources are represented by $c$ colors and the resource allocation problem is posed as a distributed vertex coloring problem of the graph $G$ using $c$ colors.

Nodes in a graph are said to be of the same \textit{type} if their neighbors have the same degree distribution. Degree distribution of a graph is defined as a vector containing the number of nodes of a given degree. The $R$-hop neighbors of a node are defined as the leaves of a depth-$R$ tree centered at the given node. A graph is defined to be \textit{$R$-hop symmetric} if there are two connected vertices with the same type of $r$-hop neighbors for all $1\leq r \leq R$. Each node is associated with two states, the \emph{color state} (color of the node) and the \emph{conflict state}. 
The color state is used interchangeably to mean the graph's color vector (the $n$ length vector representing the color of each node) or a node's color and the meaning will be clear from the context. The conflict state is defined as follows.
A node is said to be \textit{in conflict} ($C$) or \textit{not in conflict} ($\bar{C}$) if its color is same or different than \emph{all} of its neighbors respectively. 
Otherwise, it is said to be in a \textit{confused} state ($X$). 

We consider the class of $n$-user cascaded interference channel as shown in Figure~\ref{fig:model}. The graph of this network is a path graph $P_{n}$. The minimum number of colors needed to do a proper coloring of a graph is defined as the chromatic number $\chi(G)$ of the graph. The number of ways an optimal allocation can be done for $P_{n}$ with $c$ colors is given by the chromatic polynomial $\gamma_n^{(c)}(G)$. The chromatic number of a path graph $P_{n}$ is $\chi=2$ and the chromatic polynomial is $\gamma_{n}^{(c)}=c(c-1)^{n-1}$ (see, for example \cite{nb93}). There are two types of nodes in a path graph, two degree $1$ nodes at the edges, $n-2$ degree $2$ nodes in the middle.

\subsection{Protocol Description \label{sec:proto}}
In this paper, the coloring protocols used to color a graph distributedly, consist of two phases. The first phase involves starting with a random coloring and gathering information about the random starting colors of the neighboring nodes by passing messages for $R$ rounds. The second phase involves deciding whether to change color or do nothing, based on the information gathered.

First we define the message passing protocol used to gather starting state colors in the first phase. A \textit{round} of message passing is said to be complete when all nodes in the flow graph have broadcasted a message (see, for example \cite{aas10}). Each message contains the sender node's ID in the header. The message sent by node $i$ is received by all nodes $j\in V$ if  $(i,j)\in E$. During round $r$, the message of each node contains the following triplet (the ID of its $r$-hop neighbors, the initial state of its $r$-hop neighbors, parents of these $r$-hop neighbors). The parent of node is defined as the node from which the sender heard about it during the previous round. Each node $i$ maintains a local topology graph $G_{i}^{r}$ with itself as the root. After each round the new nodes are added as leaves to the corresponding branch of the local topology tree, using the parent information.

Next we define the structure of the decision function evaluated at a node $i$ after $r$ rounds of message passing. The decision function takes the local $r$-hop graph $G_{i}^{r}$ centered at node $i$ as the input and outputs the final color of the node, i.e. $f : H_{i}^{r} \to \{1,\ldots,c\}$, where $H_{i}^{r}$ is the set of all radius $=r$ graphs centered at node $i$. The decision function doesn't consider the node IDs to compute the output. For a path graph, $H_{i}^{1}$ is either a tree with one or two leaves. After one round of messaging, any node $i$ can calculate its conflict state and degree. Additionally, a degree 1 node can be in $C$ or $\bar{C}$ conflict state, while a degree 2 node can be in $C$, $\bar{C}$ or $X$ conflict state. So the inputs to the decision function in a path graph can be of five different types of local trees. Depending on whether the protocol changes the color or not for each of these five input types, there can be $32$ possible protocols. We index the protocols by the conflict states when the decision function changes the color of the node, e.g. $(C,\phi)$ refers to a protocol which changes the color of degree 1 nodes in conflict and $(C,CX)$ refers to a protocol which changes the color of a degree 1 one node in conflict or a degree 2 node in a conflict or confused state. 
%

\subsection{Performance Metric \label{sec:perf}}

For a centralized network with enough available colors ($c\geq \chi(G)$), the optimal allocation is a proper coloring with no conflicts. 
To compare the different distributed protocols, we use \emph{defects} as the performance metric, defined as the number of edges joining nodes with the same color. A defect group is defined as a connected subgraph with identically colored nodes. A defect group with $k$ nodes is denote it by $g_{k}$.
The number of defects in a given graph state is given by, 
$ \sum_{(i,j)\in E} I(i,j), $
where the identity function $I(i,j)$ is $1$ if nodes $i$ and $j$ have the same color and $0$ otherwise. The number of states that have $d$ defects are denoted by $N(d)$. The defect distribution is given by the $n$ length vector $D = [N(d)]_{d=0}^{n-1}$.
The average number of defects for a protocol is defined as,
$\bar{N}=\frac{1}{c^{n}}\sum_{d=0}^{n-1}d N(d)$, since there are $c^{n}$ different possible colorings.
Finally, we define two classes of protocols that qualify the worst case performance.
\begin{definition}[$k$-round successful coloring protocol]
A distributed coloring protocol with decision function $f_s$ is evaluated at each node of a given connected graph $G$. If the network converges to a defect free coloring after $k$-rounds of message passing, independent of any starting state, then $f_s$ is said to be $k$-round successful for $G$.
\end{definition}
\begin{definition}[$k$-round universal coloring protocol]
If there exists a distributed coloring protocol with decision function $f_u$ that is $k$-round successful for all connected network topologies then $f_u$ is said to be $k$-round universal.
\end{definition}

\section{Impossibility of an universal protocol \label{sec:results}}
Now we show that the distributed defect free coloring is impossible even for some of the simple symmetric networks.

\begin{theorem}
There exist no $R$-round successful coloring protocol for any network whose graph $G$ is $R$-hop symmetric for $R<\text{dia}(G)$.
\label{thm:conv}
\end{theorem}
\begin{IEEEproof}
If $c<\chi(G)$, the statement is trivially true, since even an optimal allocation is not defect free. For $c\geq \chi(G)$, consider two adjacent nodes $i$ and $j$ with $R$-hop symmetry and their $r$-hop ($1\leq r \leq R$) neighbors $(X_r,Y_{r-1})$ and $(Y_r,X_{r-1})$ respectively, as shown in Figure~\ref{fig:genSymm}. Consider a symmetric starting state where $X_{r}$ and $Y_{r}$, $0\leq r\leq R$ pick the same initial colors, where $X_{0}$ and $Y_{0}$ refer to nodes $i$ and $j$ respectively. After the first round, each node has information about the colors of its $r$-hop neighbors only. If there exists any protocol which works, it has to be independent of the global topology, as it is unknown to any node after the first round. 
\begin{figure}[htbp]
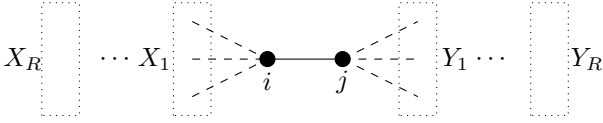

\centering
\tikz{
	\draw [fill=black] (1,1) circle (0.1cm); 
	\draw [fill=black] (2,1) circle (0.1cm); 
	\draw (1,1) -- (2,1);
	\draw [dashed] (0,1) -- (1,1);
	\draw [dashed] (0,1.5) -- (1,1);
	\draw [dashed] (0,0.5) -- (1,1);
	\draw [dashed] (2,1) -- (3,1);
	\draw [dashed] (2,1) -- (3,1.5);
	\draw [dashed] (2,1) -- (3,0.5);
	\draw (1,.7) node {$i$};
	\draw (2,.7) node {$j$};
	\draw (-.5,1) node {$X_1$};
	\draw (3.5,1) node {$Y_1$};
	\draw [dotted] (-.25,.25) rectangle (.25,1.75);
	\draw [dotted] (2.75,.25) rectangle (3.25,1.75);
	\draw (-1,1) node {$\cdots$};
	\draw (4,1) node {$\cdots$};
	\draw (-2.25,1) node {$X_R$};
	\draw (5.25,1) node {$Y_R$};
	\draw [dotted] (-2,.25) rectangle (-1.5,1.75);
	\draw [dotted] (4.5,.25) rectangle (5,1.75);
}
\caption{A symmetric subgraph with connected nodes $i$, $j$ and their $R$-hop neighbors. Note that $X_{r}$ is the $r$-hop neighbor of $i$ and the $(r+1)$-hop neighbor of $j$.
\label{fig:genSymm}}
\end{figure}

The decision function for node $i$ and $j$ are $f(G_{i}^{R})$ and $f(G_{j}^{R})$ respectively. By the symmetry of the graph, $X_r$ and $Y_r$ contain the same number and type of nodes. Additionally, by the symmetry of the starting state, the depth-$r$ leaves in  $G_{i}^{R}$ and $G_{j}^{R}$ have the same initial color for $0\leq r \leq R$. Hence the output of the decision functions are identical,  $f(G_{i}^{R})=f(G_{j}^{R})$. Hence their final colors are same and the final state of the graph cannot be defect-free. Similarly, if the protocol is randomized, optimal allocation can not be guaranteed, since there is always a non-zero probability that $i$ and $j$ will have the same final color. $R \leq \text{dia}(G)$ is assumed to keep any node from acquiring global information.
\end{IEEEproof}
\begin{corollary}
This theorem implies, there exists no $R$-round universal coloring protocol, since there is no $R$-round successful coloring protocol for any graph with a connected $R$-hop symmetric subgraph.
\end{corollary}


\section{Protocols with $1$-hop information \label{sec:achproto}}
As mentioned in Section~\ref{sec:proto}, there are $32$ different protocols possible after one round of message passing. We now consider three main protocol classes. All others exhibit similar defect performance in Monte-Carlo simulations.
We start with the random initial assignment, which is optimal when there is no information about any neighbor, i.e. before any rounds of message passing is done.
\subsection{Random initial assignment}
Since all the nodes select their colors uniformly at random, the defect distribution of a random allocation is same as the defect distribution for the set of all possible allocations with $c$ colors. 
\begin{theorem}
For a random assignment with $c$ colors, the defect distribution for a path network $P_{n}$ is given by the $n$ length vector $\mathbf{N}_{0}$, whose $d^{\text{th}}$ component is,
$$N_{0}(d) = c(c-1)^{n-d-1}{n-1 \choose d}, \text{ for } 0 \leq d \leq n-1,$$ and the average number of defects is given by,
$ \bar{N}_{0}=\frac{n-1}{c}.$
\end{theorem}
\begin{proof}
To calculate the number of states with $0\leq d \leq n-1$ defects in $P_{n}$, we start with an optimal state for $P_{n-d}$, which has no defects, introduce $d$ defects and calculate the number of ways of doing so. For each of these optimal states of $P_{n-d}$, there are $n-d$ singleton groups of colors ($g_{1}$). A defect can be introduced if any of the singleton groups is replaced by two copies of the same color. To get $d$ defects, we can choose $1\leq i \leq d$ groups and allocate some defects to each group that adds up to $d$. Since there are $n-1$ places between the $n$ nodes where the defects can be introduced, the total number of ways for arranging $d$ defects in $P_{n}$ is ${n-1 \choose d}$. Now, if we coalesce all the groups into singleton groups $g_{1}$, the number of singleton groups is $n-d$. Hence the number of ways in which these groups can be chosen from $c$ colors is same as the number of optimal states in $P_{n-d}$ which is $\gamma_{n-d}^{(c)}$.

For a path network, the chromatic polynomial is given by $\gamma_{n}^{(c)}=c(c-1)^{n-1}$. Total number of states with degree $d$ is given by, $$N_{0}(d)=\gamma_{n-d}^{(c)}{n-1 \choose d}= c(c-1)^{n-d-1}{n-1 \choose d}.$$ It is interesting to note that the defect distribution for $P_{n}$ turns out to be $\gamma_{n-d}^{(c)}$ times the $n^{\text{th}}$ row of the Pascal's triangle.
The average number of defects can now be calculated as, 
\begin{align}
\bar{N}_{0} &= \frac{1}{c^{n}}\sum\limits_{d=0}^{n-1}d\cdot c(c-1)^{n-d-1}{n-1 \choose d} \notag \\
& = \frac{(c-1)^{n-1}}{c^{n-1}}\sum\limits_{d=0}^{n-1}\frac{d}{(c-1)^d}{n-1 \choose d}.
\label{eqn:one}
\end{align}
Now, consider the expansion $\left(1+\frac{e^{x}}{\alpha}\right)^{n-1} = \sum\limits_{r=0}^{n-1} \frac{e^{rx}}{\alpha^{r}} {n-1 \choose r} $.
Differentiating both sides w.r.t. $x$, 
$$ \frac{n-1}{\alpha} \left( 1+\frac{e^{x}}{\alpha} \right)^{n-2}e^{x} = \sum\limits_{r=0}^{n-1} \frac{r}{\alpha^{r}} e^{rx} {n-1 \choose r}. $$
The right hand side of the above equation is same as the right hand side of Equation~\ref{eqn:one} for $x=0$ and $\alpha = c-1$. So, the average number of defects is given by,
$ \bar{N}_{0} 
= \frac{(c-1)^{n-1}}{c^{n-1}} \frac{(n-1)c^{n-2}}{(c-1)^{n-1}}
= \frac{n-1}{c}. $
\end{proof}

Now we calculate the length of the defect groups in all possible random assignments.
\begin{theorem}
The number of times a defect group $g_{i}$ occurs in a path network $P_{n}$ out of all random allocations using $c$ colors, is given by,
\begin{equation*}
G_i = \left\{ \begin{array}{ll} 
c, \qquad \qquad \qquad \quad~\text{ if }i=n& \mbox{} \\ 
2c(c-1), \qquad \qquad \text{ if }i=n-1& \mbox{} \\ 
c^{n-1-i}(c-1)\Big((n-i+1)c-(n-i-1)\Big), & \mbox{o.w..} 
\end{array}
\right.
\end{equation*}
\end{theorem}
\begin{proof}
The starting state can have a defect group of size $n$ in $c$ ways, one for each color. Similarly the starting state can have a group of size $n-1$ in $c$ ways and the remaining node can choose a color from the remaining $c-1$ colors in $c-1$ ways. Finally, they can be arranged in 2 ways, making the total number of ways, $2c(c-1)$. For any group of size $1 \leq i \leq n-2$, the group can either occur on one of the edges of the path network or somewhere in the middle. The color of the group can be chosen in $c$ ways. If it occurs on one of the edges, the immediate neighbor of the group can have $c-1$ possible states. The rest of the network can be selected in $c^{n-i-1}$ ways. If the defect group is in the middle, both its neighbors can choose a state in $c-1$ ways ($(c-1)^{2}$ in total). The rest of the network can be selected in $c^{n-i-2}$ ways. Now the defect group can be placed in the rest of the network in $n-1-i$ ways.  Hence the total number of ways is $2(c-1)c^{n-i-1}+(n-i-1)(c-1)^{2}c^{n-2-i}=c^{n-1-i}(c-1)\Big((n-i+1)c-(n-i-1)\Big)$.
\end{proof}
The typical occurrence of these groups are then given by the vector $\mathbf{G}=\frac{1}{c^{n}}[G_{i}]_{i=1}^{n}$. In other words, for a random allocation, a typical $P_{n}$ has $\frac{G_{i}}{c^{n}}$ defect groups of size $i$. 
These two results show that the random assignment produces $O(\frac{n}{c})$ defects typically which occur in many smaller defect groups. This observation will guide us in our one round protocols. Now we show the performance of a protocol that only corrects the errors in the edges of the network.

\subsection{Edge correcting protocol}
This refers to the protocol whose decision function is indexed by the tuple $(C,\phi)$. According to this protocol, the degree 1 nodes change their color if they are in conflict with their neighbor. Since all the path networks (except $P_{1}$) have only two degree 1 nodes, this protocol can at best decrease the number of defects by 2 at the two boundaries of the network. 

\begin{theorem}
After one round of message passing, the final defect distribution for the edge correcting protocol with $c$ colors, is given by $$N_{1}(d)=c^{3}(c-1)^{n-d-3}{n-3 \choose d}, \text{ for } 0 \leq d \leq n-3,$$ and the average number of defects is given by, $\bar{N}_{1}=\frac{n-3}{c}$.
\end{theorem}

\begin{proof}
As shown in Figure~\ref{fig:protocol1}, the center of the network containing $n-2$ nodes remain unchanged. For any starting state ($a,x,y,b \in \{1,\ldots,c\}$) the two degree 1 nodes end up in state $\bar{x} \in \{1,\ldots,c\}\diagdown x$ and $\bar{y} \in \{1,\ldots,c\}\diagdown y$ respectively.

\begin{figure}[htbp]
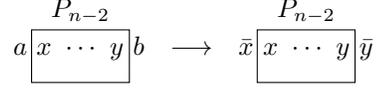

\centering
\tikz{
	\draw (1,1) node {$a~x~\cdots~y~b~~\longrightarrow~~\bar{x}~x~\cdots~y~\bar{y}$};
	\draw (-1.15,0.55) rectangle (0.15,1.25);
	\draw (-0.5,1.5) node {$P_{n-2}$};
	\draw (2.5,1.5) node {$P_{n-2}$};
	\draw (1.85,0.55) rectangle (3.15,1.25);
}
\caption{ Edge correcting protocol can reduce at most two defects at the edges of the path graph.
\label{fig:protocol1}}
\end{figure}

Since the edges on the boundary do not have any defects, the number of defects in the final state is the same as the number of defects in the fixed part as shown in Figure~\ref{fig:protocol1}. Since all the $c^{2}$ starting states with the same fixed part (corresponding to the $c^{2}$ possible values of the first and last nodes) converge to the same final state as shown in the figure, the defect distribution of the final state for $P_{n}$ is same as the $c^{2}$ times the defect distribution of $P_{n-2}$. Hence $$N_{1}(d) = c^{2}N_{d}(P_{n-2})= c^{3}(c-1)^{n-d-3}{n-3 \choose d}.$$
Similarly, the average number of defects is given by,
\begin{align*}
\bar{N}_{1} &= \frac{1}{c^{n}}\sum\limits_{d=0}^{n-1}d\cdot c^{3}(c-1)^{n-d-3}{n-3 \choose d} \\
&= \frac{(c-1)^{n-3}}{c^{n-3}} \frac{(n-3)c^{n-4}}{(c-1)^{n-3}}
= \frac{n-3}{c}.
\end{align*}
\end{proof}
Even though the edge correcting protocol corrects at most two defects in the whole network, the average number of defects produced by other protocols, that correct the edges of each group, closely match its performance when $n$ is large in Monte-Carlo simulations. 
\subsection{Center correcting protocol}
This refers to the protocol whose decision function is indexed by the tuple $(\phi,C)$. According to this protocol, only the degree 2 nodes change their assigned color if they are in conflict. 
The center of each group changes color and the boundaries remain fixed, as shown in Figure~\ref{fig:protocol2}. A node with initial color $x$ changes randomly to a color from $\{1,\ldots,c\} \setminus x$. If $c>2$, this protocol can correct more than or equal to two defects for all groups except $g_{2}$. 
Hence each group $g_{n_{i}}$ of size $n_{i}$, can give rise to a defect distribution of $D_{i} = [N_{0}(d;P_{n_{i}-2},c-1)]_{d =0}^{n_{i}-3}$. If $P_{n}$ has $i$ groups $g_{n_{1}},\ldots,g_{n_{i}}$, the defect distribution $D$ after this protocol is given by the convolution of the corresponding defect distributions for the random allocations of groups $g_{n_{1}-2},\ldots,g_{n_{i}-2}$, i.e. $D = \gamma_{n_{i}}^{(c)}D_{1}\otimes \cdots \otimes D_{i}$. Hence, the number of states with defect $d$ in the final defect distribution is given by,
$$ D(d) = \sum_{0\leq i_{j} \leq d, \sum_{j} i_{j} = d}\gamma_{n_{i}}^{(c)} N_{j}(P_{i_{j}},c-1).$$
For $c=2$, this can be evaluated in closed form. 

\begin{theorem}
After one round of message passing, the final defect distribution for the center correcting protocol with $c=2$ colors, is given by 
\small 
\begin{equation*}
N_{2}(d)=2\sum_{k=0}^{d}\sum_{i=k+1}^{K_{k}^{*}} {n-d_{s} \choose i} {i \choose k}{d_{s}+i-2k-1 \choose i-k-1} +2 {n-d_{f} \choose d}.
\end{equation*}
\normalsize
\end{theorem}

\begin{proof}
The aim is to find the number of states that end up with $d$ defects after applying the center correcting protocol to a random initialization with two colors. Note that the presence of $g_{1}$ does not result in any defects in the starting or the final states. On the other hand the protocol has no effect on $g_{2}$. For all other groups, the center correcting protocol reduces the number of defects by two, since $g_{k}$ gets transformed to $g_{k-2}$. 

\begin{figure}[htbp]
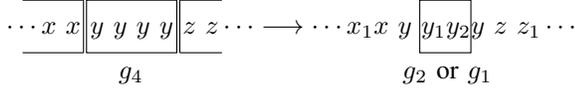

\centering
\tikz{
	\draw (1,1) node {$\cdots x~x~y~y~y~y~z~z \cdots \longrightarrow \cdots x_{1}x~y~y_{1} y_{2}y~z~z_{1} \cdots$};
	\draw (-2.6,0.7) -- (-1.79,0.7);
	\draw (-1.79,0.7) -- (-1.79,1.4);
	\draw (-1.79,1.4) -- (-2.6,1.4);
	\draw (-1.75,0.7) rectangle (-0.55,1.4);
	\draw (0.05,1.4) -- (-0.51,1.4);
	\draw (-0.51,1.4) -- (-0.51,0.7);
	\draw (-0.51,0.7) -- (0.05,0.7);
	%
	\draw (-1.16,0.4) node {$g_{4}$};
	\draw (2.68,0.7) rectangle (3.36,1.4);
	\draw (3.05,0.4) node {$g_{2}$ or $g_{1}$};
	%
}
\caption{ Center correcting protocol can reduce at least two defects for each group (except $g_{2}$) in the path graph. Notationwise, $x_{1}\in \{1,\ldots,c\} \setminus x$ etc. 
\label{fig:protocol2}}
\end{figure}

If there are $i$ groups in the starting state and $k$ of them are $g_{2}$, in order to have $d$ defects in the final state, the starting state needs to have $d_{s}=d+2(i-k)$ defects. So, we start with $P_{n-d_{s}}$ with an optimal allocation and introduce $d_{s}$ defects in $i$ places with $k$ number of $g_{2}$ groups. The $i$ places can be chosen in ${n-d_{s} \choose i}$ ways. Out of these $i$ places, the $k$ places for $g_{2}$ groups can be chosen in ${i \choose k}$ ways. Now, the number of ways to introduce the remaining $d_{s}-k$ defects in $i-k$ places is same as (written in terms of generating functions)
\begin{align*}
&=\text{coefficient of }x^{d_{s}-k} \text{ in } (x^{2}+x^{3}+\cdots)^{i-k}\\
&=\text{coefficient of }x^{d_{s}-k-2(i-k)} \text{ in } (1-x)^{-(i-k)}\\
&= {d+i-2k-1 \choose i-k-1}.
\end{align*}
Hence, the number ways we can get $d$ defects in the final state such that there are less than $d$ number of $g_{2}$ groups in the starting state is given by $\sum_{k=0}^{d}\sum_{i=k+1}^{K_{k}^{*}}\gamma_{n-d_{s}}^{(2)}{n-d_{s} \choose i} {i \choose k}{d+i-2k-1 \choose i-k-1}$, where $K_{k}^{*}=\lfloor \frac{n-1-d+2k}{2} \rfloor$.

Finally, the number of ways to have $d$ number of $g_{2}$ groups and the rest $g_{1}$ groups is given by $\gamma_{n-d}^{(2)} {n-d \choose d}$. Hence the total number of states with $d$ defects is given by,
$2\sum_{k=0}^{d}\sum_{i=k+1}^{K_{k}^{*}} {n-d_{s} \choose i} {i \choose k}{d_{s}+i-1 \choose i-k-1} +2 {n-d \choose d}.$
\end{proof}

\begin{figure}[ht]
\centering
\epsfxsize=3in
\epsffile{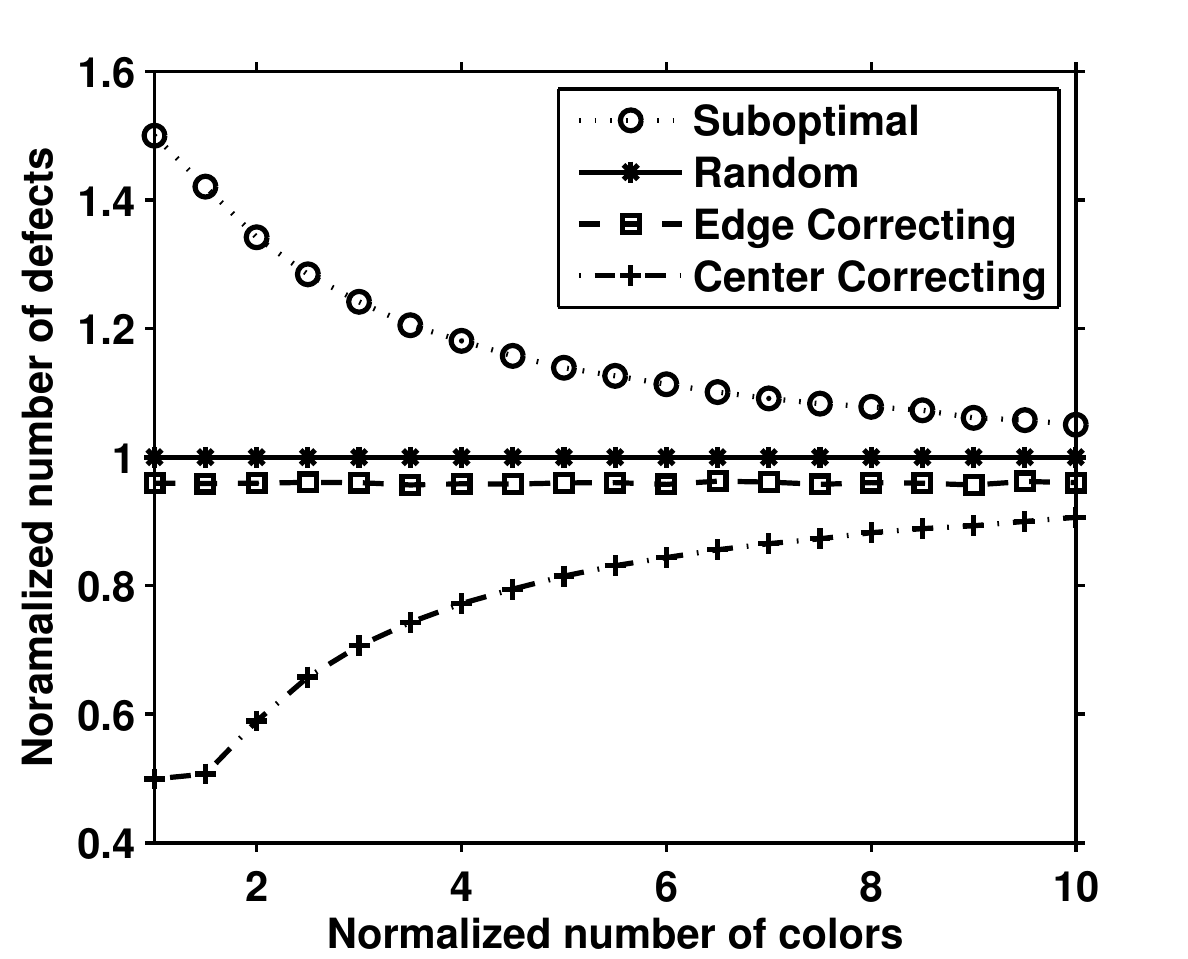}
\caption{\it Defects as a function of number of colors available for $P_{50}$. Defects for each protocol are normalized with the number of defects for a random assignment and the number of available colors is normalized by the chromatic number $\chi(P_{50})=2$. The suboptimal curve refers to the $(\bar{C},\bar{C}X)$ protocol.
\label{fig:result2}}
\vspace{-.5cm}
\end{figure}

Hence the center correcting protocol does random assignment (by leaving out one color) in each small defect group created by the initial random assignment. As shown by the Monte-Carlo results in Figure~\ref{fig:result2} both the edge and center correcting protocols perform close to the random assignment when the number of colors is large. There are other protocols like $(\bar{C}, \bar{C}X)$ which sometimes increase the number of defects, perform worse than a random assignment. However even these protocols perform very close to a random assignment for large $c$. All other possible protocols cluster around these three classes, but are not shown for clarity of the plot.

\section{Conclusion \label{sec:conclusion}}
In summary, we show that for even the simple case of $R$-hop symmetric networks, there can be no $R$-round successful message passing protocol and in general there can be no $R$-round universal protocol. Even though distributed decision result in a performance loss due to the existence of defects, we show that this can be offset by having more colors than the chromatic number of the graph. We show that the best protocol after one round of message passing does another round of random allocation in each defect group. Finally we show through simulations that the loss due to distributed decisions decreases with an increase in the number of available colors.

\bibliographystyle{IEEEtranS}
\bibliography{references_dash}

\begin{thebibliography}{10}
\providecommand{\url}[1]{#1}
\csname url@samestyle\endcsname
\providecommand{\newblock}{\relax}
\providecommand{\bibinfo}[2]{#2}
\providecommand{\BIBentrySTDinterwordspacing}{\spaceskip=0pt\relax}
\providecommand{\BIBentryALTinterwordstretchfactor}{4}
\providecommand{\BIBentryALTinterwordspacing}{\spaceskip=\fontdimen2\font plus
\BIBentryALTinterwordstretchfactor\fontdimen3\font minus
  \fontdimen4\font\relax}
\providecommand{\BIBforeignlanguage}[2]{{%
\expandafter\ifx\csname l@#1\endcsname\relax
\typeout{** WARNING: IEEEtranS.bst: No hyphenation pattern has been}%
\typeout{** loaded for the language `#1'. Using the pattern for}%
\typeout{** the default language instead.}%
\else
\language=\csname l@#1\endcsname
\fi
#2}}
\providecommand{\BIBdecl}{\relax}
\BIBdecl

\bibitem{aas10}
V.~Aggarwal, S.~Avestimehr, and A.~Sabharwal, ``How (information theoretically)
  optimal are distributed decisions?'' in \emph{44th Annual conference on
  information sciences and systems}, 2010.

\bibitem{bgs98}
M.~Bellare, O.~Goldreich, and M.~Sudan, ``Free bits, {PCP}s and
  non-approximability - towards tight results,'' \emph{{SIAM} Journal on
  Computing}, vol.~27, pp. 804--915, 1998.

\bibitem{nb93}
N.~Biggs, \emph{Algebraic graph theory}, 2nd~ed.\hskip 1em plus 0.5em minus
  0.4em\relax Cambridge University Press, 1993.

\bibitem{bgp06}
S.~Boyd, A.~Ghosh, B.~Prabhakar, and D.~Shah, ``Randomized gossip algorithms,''
  \emph{{IEEE} Trans. Inf. Theory}, vol.~52, no.~6, pp. 2508--2530, jun 2006.

\bibitem{ca06}
C.~M. Cordeiro and D.~P. Agarwal, \emph{Ad hoc and sensor networks - theory and
  applications}.\hskip 1em plus 0.5em minus 0.4em\relax World Science
  Publication Company, 2006.

\bibitem{fk98}
U.~Feige and J.~Kilian, ``Zero knowledge and the chromatic number,''
  \emph{Journal of Computer and System Sciences}, vol.~57, pp. 187--199, 1998.

\bibitem{fgi09}
P.~Fraigniaud, C.~Gavoille, D.~Ilcinkas, and A.~Pelc, ``Distributed computing
  with advice: information sensitivity of graph coloring,'' \emph{Distributed
  Computing}, vol.~21, pp. 395--403, 2009.

\bibitem{kar72}
R.~M. Karp, ``Reducibility among combinatorial problems,'' in \emph{Proceedings
  of Symposium on Complexity of Computer Computations}, 1972, pp. 85--103.

\bibitem{kos06}
K.~Kothapalli, M.~Onus, C.~Scheideler, and C.~Schindelhauer, ``Distributed
  coloring in $o(\log n)$ bit rounds,'' in \emph{20th IEEE Inernational
  Parallel and Distributede Processing Symposium (IPDPS)}, 2006.

\bibitem{kk02}
M.~Kubale and L.~Kuszner, ``A better practical algorithm for distributed graph
  coloring,'' in \emph{International conference on parallel computing in
  electrical engineering, PARELEC}, 2002, p.~72.

\bibitem{lin92}
N.~Linial, ``Locality in distributed graph algorithms,'' \emph{{SIAM} Journal
  on Computing}, vol.~21, no.~1, pp. 193--201, 1992.

\bibitem{mbk09}
\BIBentryALTinterwordspacing
K.~Moshksar, A.~Bayesteh, and A.~K. Khandani. (2009) A model for randomized
  resource allocation in decentralized wireless networks. [Online]. Available:
  \url{http://arxiv.org/abs/0911.5527}
\BIBentrySTDinterwordspacing

\bibitem{pel00}
D.~Peleg, \emph{Distributed computing: {A} locality-sensitive approach}.\hskip
  1em plus 0.5em minus 0.4em\relax SIAM, 2000.

\bibitem{ram99}
S.~Ramanathan, ``A unified framework and algorithm for channel assignment in
  wireless networks,'' \emph{Wireless Networks}, vol.~5, pp. 81--94, 1999.

\bibitem{xln06}
Y.~Xue, B.~Li, and K.~Nahrstedt, ``Optimal resource allocation in wireless ad
  hoc networks: A price-based approach,'' \emph{{IEEE} Transactions on Mobile
  Computing}, vol.~5, apr 2006.

\end{thebibliography}
\end{document}